\documentclass[floatfix,
reprint,
superscriptaddress,
nofootinbib,
aps, pra
]{revtex4-1}

\usepackage{custom}
\pdfoutput=1
\begin{document}

%========================================
\title{Finite resolution ancilla-assisted measurements of quantum work distributions}

\author{Shadi Ali Ahmad}
\email[]{shadi.ali.ahmad.22@dartmouth.edu}
\affiliation{Department of Physics and Astronomy, Dartmouth College, Hanover, New Hampshire 03755, USA}

\author{Alexander R. H. Smith}
\email[]{arhsmith@anselm.edu}
\affiliation{Department of Physics, Saint Anselm College, Manchester, New Hampshire 03102, USA}
\affiliation{Department of Physics and Astronomy, Dartmouth College, Hanover, New Hampshire 03755, USA}

\date{\today}

\begin{abstract}
Work is an observable quantity associated with a process, however there is no Hermitian operator associated with its measurement. We consider an ancilla-assisted protocol measuring the work done on a quantum system driven by a time-dependent Hamiltonian via two von-Neumann measurements of the system's energy carried out by a measuring apparatus modeled as a free particle of finite localization and interaction time with the system. We consider system Hamiltonians which both commute and do not commute at different times, finding corrections to fluctuation relations like the Jarzynski equality and the Crooks relation. This measurement model allows us to quantify the effect that measuring has on the estimated work distribution, and associated average work done on the system and average heat exchanged with the measuring apparatus.
\end{abstract}

\maketitle

%========================================

%========================================
%========================================
\label{Introduction}
Central to thermodynamics is the notion of work, which is defined classically as a line integral over a definite trajectory through the configuration space associated with the system of interest. However, no such trajectory exists for a quantum system. Work is not a state function, but instead depends on the process under which a system evolves, and it is for this reason that work is not represented by a Hermitian operator~\cite{talknerFluctuationTheoremsWork2007,campisiColloquiumQuantumFluctuation2011,talknerAspectsQWork2016} (however, see Ref.~\cite{silvaqwork2021} for such a proposal). For a closed system, which does not exchange heat with its environment, it follows from the first law that the work done on the system is equal to the change in its energy. This suggests that we measure the energy of the system twice, once at the beginning and once at the end of a process, and attribute the difference in the outcomes of these measurements to the average work done on the system.

It was initially thought that these two energy measurements must be performed via two projective measurements of the system directly~\cite{talknerFluctuationTheoremsWork2007,campisiColloquiumQuantumFluctuation2011}. However, Roncaglia \emph{et al.}~\cite{roncagliaWorkMeasurementGeneralized2014} later realized that work may be measured via a von Neumann measurement model that involves a measuring apparatus that interacts with the system twice, after which a single projective measurement of the apparatus is made and the outcome is associated with the work done on the system.
In contrast to performing two projective energy measurements, the latter approach has the advantage of not destroying coherence among energy eigenstates of the system, thus allowing for the investigation of quantum coherence on thermodynamic processes~\cite{lostaglioQuantumCoherenceTimeTranslation2015a,lostaglioDescriptionQuantumCoherence2015, gherardiniEndPointMeasurement2021}. In addition, because only one projective measurement needs to be made to obtain a value of work, the experimental implementation of this method for measuring work may be easier~\cite{chiaraMeasuringWorkHeat2015a}.

Any realistic measurement scheme will be of finite duration and the free evolution of the measuring apparatus may affect the measured work distribution~due to the non-ideality of the measurements~\cite{debarba2019work}. While weak measurements may be used to model these finite resolution effects~\cite{PhysRevE.92.042150}, we are interested in situations where the measurement process can significantly affect the behavior of the system on account of short-time scale measurement interactions. It is the purpose of this article to explore the consequences of such realistic measurements on the scheme proposed by Roncaglia \emph{et al.}~\cite{roncagliaWorkMeasurementGeneralized2014}.

We begin in Sec.~\ref{Measuring work} by describing a measurement model for the work done by a time-dependent Hamiltonian that takes into account the finite duration of the interactions between the system of interest and measuring apparatus and the free evolution of the measuring apparatus during the process that is performing work; related measurement models have been examined in the past~\cite{talknerAspectsQWork2016, solinasvariance2017}. In Sec.~\ref{thermocons} we establish a method to estimate the effect of the non-ideality of the measurement model on thermodynamic quantities like the average work and heat. In Secs.~\ref{Self-commuting system Hamiltonians}
and~\ref{not self-commuting}, this measurement model is applied to processes described by time-dependent Hamiltonians that respectively do and do not self-commute at different times. In doing so, we derive modifications to the Crooks relation and Jarzynski equality~\cite{crooksEntropyProductionFluctuation1999,JarzynskiEquality1996} stemming from the finite duration of these measurements and give a physical interpretation of the modifications in terms of heat exchange between the measuring apparatus and system. We also consider estimates of the average heat flow between system and measuring apparatus, where in the self-commuting case this vanishes while in the non-self-commuting case it is generically non-zero. Moreover, in Sec.~\ref{not self-commuting} we analyze the effects of the parameters of the measurement model on the sampled work distributions. We summarize our results in Sec.~\ref{Conclusion}.

Throughout we will work with $\hbar=1$. Further, $\mathcal{S}(\mathcal{H})$ and $\mathcal{E}(\mathcal{H})$ will denote respectively the space of density operators\footnote{$\mathcal{S}(\mathcal{H}) \ce \{  \rho \in \mathcal{T}(\mathcal{H})  \ | \ \rho \succeq 0 \mbox{ and } \tr \rho = 1\}$, where $\mathcal{T}(\mathcal{H})$ is the space of trace class operators acting on $\mathcal{H}$.} and the space of effect operators\footnote{$\mathcal{E}(\mathcal{H}) \ce \{  E \in \mathcal{B}(\mathcal{H})  \ | \ 0 \preceq E \preceq I \mbox{ and } E = E^\dagger \}$, where $\mathcal{B}(\mathcal{H})$ is the space of bounded operators acting on $\mathcal{H}$.} acting on the Hilbert space~$\mathcal{H}$.
%

%========================================
%========================================
\section{Measuring work}
\label{Measuring work}

Consider a closed system that does not interact with its surroundings, so that no heat can be added to the system, described by the Hilbert space $\mathcal{H}_S$. In accordance with the first law, the work done on the system by a time-dependent Hamiltonian $H_S(t)$ between an initial time $t=t_i$ and final time $t=t_f$ is equal to the change in its internal energy 
\begin{align}
W = E_n(t_f) - E_m(t_i), \label{work}
\end{align}
where $E_n(t)$ is an eigenvalue of the system Hamiltonian associated with the eigenvector $\ket{E_n(t)} $ at time $t$, that is, $H_S(t) \ket{E_n(t)} = E_n(t) \ket{E_n(t)}$. For simplicity, we have assumed that the spectrum of $H_S(t)$ is non-degenerate and discrete (labeled by the index $n$), however, the results that follow are expected to generalize straightforwardly.

\subsection{Two-point measurement scheme}
\label{twopoint}
One of the most common operational definitions of work is the so-called two-point measurement scheme~\cite{campisiColloquiumQuantumFluctuation2011,talknerFluctuationTheoremsWork2007}. Suppose the system is prepared in the state $\rho_S(t_i) \in \mathcal{S}\left(\mathcal{H}_S\right)$. A projective measurement of the system's energy is made at $t=t_i$ yielding the outcome $E_m(t_i)$. The system then evolves  from $t_i$ to $t_f$ as described by the unitary $U_S(t_f)$ generated by $H_S(t)$. Then, the system energy is measured again yielding the outcome $E_n(t_f)$. From the outcomes of these two measurements the work performed in this particular realization of the protocol is given by Eq.~\eqref{work}. The outcomes of these energy measurements are probabilistic and thus so too is the amount of work $W$ done on the system. The probability associated with an amount of work $W$ is
\begin{align}
\mathcal{P}(W) = \sum_{m,n} \mathcal{P}_m \mathcal{P}_{m \to n} \delta \big(W - [E_n(t_f)- E_m(t_i)] \big),
\label{TMP}
\end{align}
where $\delta$ is the Dirac delta function, $\mathcal{P}_m \ce \braket{E_m(t_i) | \rho_S(t_i) | E_m(t_i)}$ is the probability of outcome $m$ in the first measurement, and  $\mathcal{P}_{m \to n} \ce \abs{\braket{E_n(t_f) | U_S(t) | E_m(t_i)}}^2$ is the probability of outcome $n$ in the second measurement conditioned on outcome $m$ in the first measurement; see Ref.~\cite{dechiaraAncillaAssistedMeasurementQuantum2018} for a recent discussion.

\subsection{Ancilla-assisted protocol}

Alternative to the two-point measurement scheme, one can consider an explicit measurement model that describes an apparatus which couples to the system at the times $t_i$ and $t_f$ in such a manner that a subsequent projective measurement of the apparatus yields the amount of work performed on the system between $t_i$ and $t_f$.

\begin{figure}[t]
\includegraphics[width= 245pt]{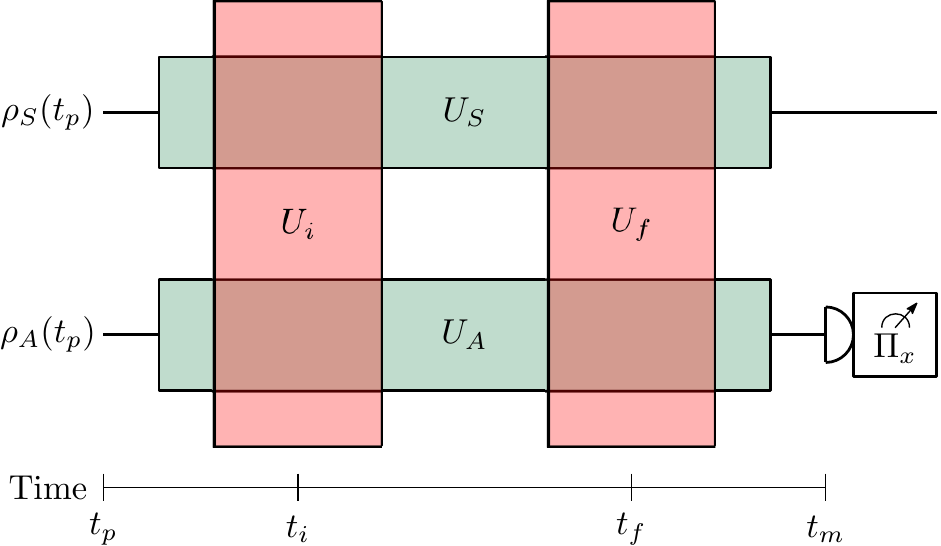}
\caption{The measurement model described in the text is depicted as quantum circuit that implements a single  measurement described by a POVM. The different evolution channels are: $U_{S}$ for the evolution of $\rho_{S}(t_{p})$ under $H_{S}$, similarly for the measuring apparatus $A$, and $U_{i},U_{f}$ represent the evolution of the joint state under the interaction Hamiltonian. 
}
\label{protocolFig}
\end{figure}

Let the measuring apparatus be modeled as a free particle on the real line, whose associated Hilbert space is $\mathcal{H}_A \simeq L^2(\mathbb{R})$ and whose free evolution is governed by the Hamiltonian $H_A = P^2/2m$, where $m$ is a mass parameter that governs the dispersion of the measuring apparatus in position space. Suppose that the system and apparatus are prepared at the time $t_p < t_i$ in the separable state $\rho_S(t_p) \otimes \rho_A(t_p)$, where  $\rho_S(t_p) \in \mathcal{S}\left(\mathcal{H}_S\right)$ and $\rho_A(t_p) \in \mathcal{S}\left(\mathcal{H}_A\right)$. For simplicity we will suppose that the apparatus is initially a pure state $\rho_A(t_p) = \ket{\psi_A(t_p)}\! \bra{\psi_A(t_p)}$ localized in position space around $x = 0$,
\begin{align}
\ket{\psi_A(t_p)} = \frac{1}{\pi^{1/4}\sqrt{\sigma_{x}}} \int dx \, e^{-\frac{x^2}{2\sigma_{x}^2}} \ket{x}, 
\label{Astatepgaus}
\end{align}
where $\ket{x}$ is the generalized eigenvector of the position operator $X$, that is, $X \ket{x} = x \ket{x}$ for all $x \in \mathbb{R}$. The apparatus must interact with the system such that it keeps a coherent record of the energy of the system at the times $t_i$ and $t_f$. An interaction Hamiltonian that accomplishes this is
\begin{align}
H_{SA}(t) =  f(t) H_S(t) \otimes \lambda P, \nn
\end{align}
where $\lambda \in \mathbb{R}$ has units of inverse energy momentum and is interpreted as the conversion factor between the displacement of the apparatus and the work done on the system, and $P$ is the momentum operator acting on $\mathcal{H}_A$,  $f(t) \ce g(t-t_f) - g(t-t_i)$, and $g(t)$ is a function with narrow support around $t=0$. Because the momentum operator $P$ generates a translation of the position operator $X$, the evolution generated by $H_S(t)$ first translates the apparatus to the left by an amount conditioned on the internal energy of the system at time $t_i$ and then translates the  apparatus to the right conditioned on the internal energy of the system at time $t_f$. The system and apparatus evolve from the time $t_p$ to $t_m>t_f$ according to the unitary operator  
\begin{align}
U_{t_p \to t_m} = \mathcal{T} e^{-i \int_{t_p}^{t_m} \dif{t} \, H(t)},  \nn
\end{align}
where $\mathcal{T}$ denotes the time ordering operator and the total Hamiltonian $H(t)$ describing the system, apparatus, and their interaction is
\begin{align}
H(t) = H_S(t) + H_A + H_{SA}(t)
\label{totalH}
\end{align}
At the time $t_m$, a position measurement of the apparatus is made. The outcome of which corresponds to the measured work $W$ in this realization of the process governed by $H_S(t)$. Accordingly, the probability density of an amount of work $W$ being done on the system is given by
\begin{align}
\mathcal{P}(W) = \tr \left[ I_S \otimes \Pi_{x = W} U_{t_p \to t_m} \rho_S(t_p) \otimes \rho_A(t_p) U_{t_p \to t_m}^\dagger \right], \nn
\end{align}
where $\Pi_{x} \ce \ket{x}\!\bra{x} \in \mathcal{E} \left(\mathcal{H}_A\right)$ is the effect operator associated with outcome $x \in \mathbb{R}$. This protocol constitutes a measurement model and is depicted in Fig.~\ref{protocolFig} as a quantum circuit.

The above measurement model induces a positive operator-valued measure (POVM) described by effect operators $ E(W) \in \mathcal{E}(\mathcal{H}_S)$ for all $W \in \mathbb{R}$ such that
\begin{align}
\mathcal{P}(W) &= \tr \big[E(W) \rho_S(t_p) \big] \nn \\
& = \tr \left[ I_S \otimes \Pi_{x = W} U_{t_p \to t_m} \rho_S(t_p) \otimes \rho_A(t_p) U_{t_p \to t_m}^\dagger \right]\!, \label{ProbabilityReproducability}
\end{align}
where the last equality defines $E(W)$ and is known as the the probability reproducibility condition~\cite{heinosaariMathematicalLanguageQuantum2011}. Inverting Eq.~\eqref{ProbabilityReproducability} allows for the POVM elements to be solved for explicitly
\begin{align}
E(W) = \bra{\psi_A(t_p)} U_{t_p \to t_m}^\dagger I_S \otimes \Pi_{x=W}  U_{t_p \to t_m} \ket{\psi_A(t_p)}.
\label{Effect}
\end{align}

In the ideal limit where the initial state of the apparatus is completely localized in the position/measurement basis, $\sigma_x \to 0$, the measurement interaction happens infinitely fast, $g(t) \to \delta(t)$, and the initial state of the system is prepared in the state $\ket{E_m(t_p)}$ with probability $\mathcal{P}_m \ce \braket{E_m(t_p) | \rho_S(t_p) | E_m(t_p)}$, then the probability distribution in Eq.~\eqref{ProbabilityReproducability} is equivalent to the work distribution sampled in the two-point measurement scheme and given in Eq.~\eqref{TMP}. Henceforth, we will refer to this limit as the \textit{ideal measurement limit}.
\label{secideal}

%========================================
%========================================
\subsection{Thermodynamic considerations}
\label{thermocons}
Suppose the system of interest is subject to a time-dependent Hamiltonian $H_S(t)$ and evolves as $\rho_S(t)$. The first law of thermodynamics states that
\begin{align}
\Braket{\Delta U} &= \Braket{W} + \Braket{Q},
\label{firstlaw}
\end{align}
where $\braket{W} \ce \int_{t_{p}}^{t_{m}} \dif{t} \, \tr{\left[ \dot{H}_{S}(t)\rho_{S}(t)\right]}$ and $\braket{Q} \ce \int_{t_{p}}^{t_{m}} \dif{t} \, \tr{\left[H_{S} (t)\dot{\rho}_{S}(t)\right]}$; see for example~\cite{Vinjanampathy2016QuantumT}.

The fact that the measurement apparatus has to interact with the system in order to sample the work distribution leads to the possibility of the apparatus performing work on the system and modifying the work distribution. Although this does not occur when using ideal von-Neumann measurements in the two-point measurement scheme \cite{debarba2019work}, we expect a different outcome based on the finite resolution of our measurement model. To examine this further we have to specify additional layers of detail defining the measurement process. The first layer involves completely ignoring the effect of the measurement interaction between the system and apparatus on the evolution of the system, $\rho_S(t)$, and corresponds to the ideal measurement limit. An additional layer of detail takes into account the measurement interaction, which in turn modifies the evolution of the system state to $\tilde{\rho}_S(t) \neq \rho_S(t) $. 
As a consequence, this results in different amounts of average work being performed on the system 
\begin{align}
    \braket{W}_{S} &\ce \int_{t_{i}}^{t_{f}} \dif{t} \tr{\left[ \dot{H}_{S}(t)\rho_{S}(t)\right]}, \label{sswork}\\
    \braket{\tilde{W}}_{S} &\ce \int_{t_{i}}^{t_{f}} \dif{t} \tr{\left[ \dot{H}_{S}(t)\tilde{\rho}_{S}(t)\right]}.
    \label{Defworktilde} 
\end{align}
The difference between these quantities,
\begin{align}
    \Delta W_{\rm int} \ce \braket{\tilde{W}}_{S}  - \braket{W}_{S},  \label{workint}
\end{align}
quantifies the additional work done on the system due to its interaction with the measuring apparatus. Using the first law in Eq.~\eqref{firstlaw}, we similarly define 
\begin{align}
    \braket{Q}_{S} &\ce \braket{\Delta U} - \braket{W}_{S}, \nn \\
    \braket{\tilde{Q}}_{S} &\ce \braket{\tilde{\Delta U}} - \braket{\tilde{W}}_{S}, \nn
\end{align}
and their difference
\begin{align}
    \Delta Q_{\rm int} \ce \braket{\tilde{Q}}_{S}  - \braket{Q}_{S}.  \label{heatint}
\end{align}

Both of the above average work quantities reference observables that are to be measured on the system itself, as opposed to an observable on the measuring apparatus. The average work computed from the work distribution is
\begin{equation}
    \braket{W}_{\rm dist}  \ce \int \dif{W} \, W \mathcal{P}(W,t_{m}). \label{povmavgwork}
\end{equation}
Similarly, the difference in the average work arising from sampling this work distribution,
\begin{align}
    \Delta W_{\rm POVM} \ce \braket{W}_{\rm dist} - \braket{W}_{S}, \label{davgworkpovm}
\end{align}
 quantifies the effect of using the measured work distribution $\mathcal{P}(W,t_{m})$ and the additional work done on the system relative to the ideal measurement limit. Note that we do not define similar quantities for the heat since that would require a prescription for calculating $\braket{\Delta U}$ using the measuring apparatus.

In Sec.~\ref{Self-commuting system Hamiltonians}, we show that $\Delta W_{\rm POVM}$, which is non-zero in general, vanishes upon taking the ideal measurement limit. More surprisingly, we find that $\Delta W_{\rm int}$ vanishes when the system Hamiltonian commutes with itself at different times, which means that the second layer of detail in describing realistic work measurements does not suffice in finding the average work imparted by the apparatus in the sense defined above. Finally, in Sec.~\ref{not self-commuting}, which is the example of a non-self-commuting Hamiltonian, we find that in general $\Delta W_{\rm POVM}$ and $\Delta W_{\rm int}$ are non-zero and differ from each other.

\begin{figure}[t]
\includegraphics[width= 245pt]{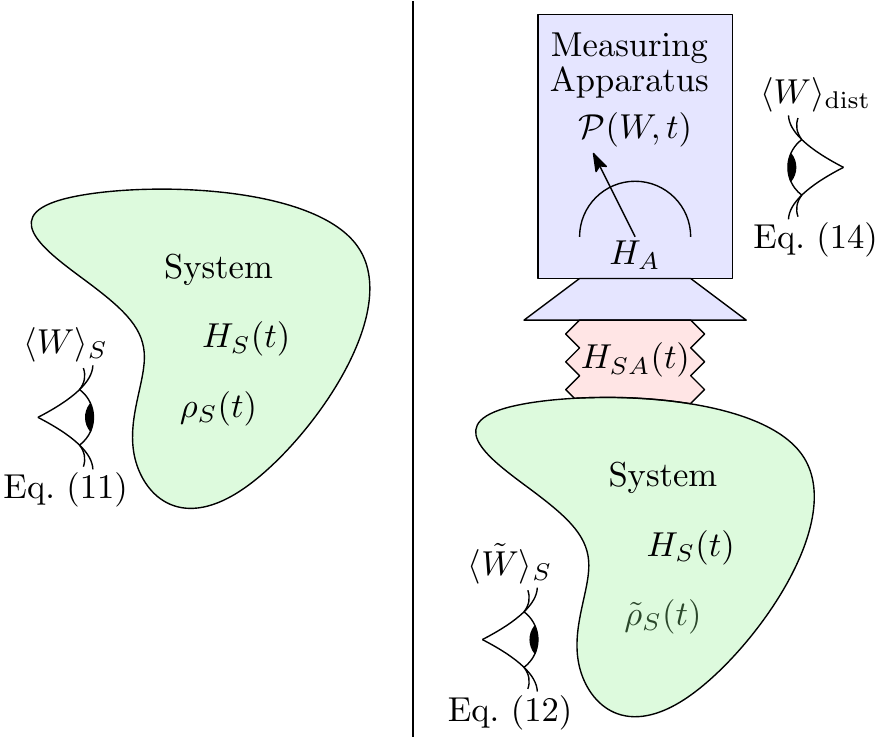}
\caption{This figure depicts the three measurement model layers outlined in Sec.~\ref{thermocons}, illustrating qualitatively their associated averages, $\braket{W}_S$, $\braket{\tilde{W}}_S$, and $\braket{W}_{\rm dist}$. The latter two definitions reduce to the former upon taking the ideal measurement limit.}
\label{protocolFiglayers}
\end{figure}

%========================================
%========================================
\section{Self-commuting System Hamiltonians}
\label{Self-commuting system Hamiltonians}
In this section, analytic expressions of the work distribution $\mathcal{P}(W,t)$ are derived using the POVM construction above for the case of systems driven by a time-dependent Hamiltonian $H_{S}(t)$ which commutes with itself at different times, $[H_S(t),H_S(t')] = 0$. This includes quantum adiabatic processes for which no heat is added to the system by $H_{S}$. For such systems, it is shown that $\Delta W_{\rm int} =0 $ and $\Delta W_{\rm POVM}$ is a function of the measurement interaction that vanishes in the ideal measurement limit discussed above. The measured work distribution is modified on account of the system-apparatus interaction, which in turn leads to corrections to both the Crooks relation and the Jarzynski equality.

\subsection{Setup}
\label{subsectionselfcommutingtheory}
Consider a system described by the Hilbert space $\mathcal{H}_S$ and whose evolution is governed  by the time-dependent Hamiltonian $H_S(t) \in \mathcal{B}(\mathcal{H}_S)$. Suppose that $H_S(t)$ commutes with itself at different times
\begin{align}
\left[ H_S(t) , H_S(t') \right] = 0, \quad \forall \ t,t' \in \mathbb{R}^{+}.
 \nn
\end{align}
It follows that for such Hamiltonians, the energy eigenbasis does not change in time, and therefore by the spectral theorem\footnote{For simplicity, we consider here the case when the spectrum $\sigma_S$ is discrete; however, the results presented here naturally generalize to the case of continuous and degenerate spectrum Hamiltonians.} 
\begin{align}
H_S(t) = \sum_{n} E_n(t) \ket{E_n}\! \bra{E_n}, \nn
\end{align}
where $\ket{E_{n}} \in \mathcal{H}_S$ is the energy eigenstate associated with the eigenvalue $E_{n}(t)\in \spec (H_S(t))$, that is, $H_S(t) \ket{E_{n}} = E_{n}(t) \ket{E_{n}}$. Thus, it is only the spectrum of the system Hamiltonian that changes in time, not its eigenbasis. An example of such a self-commuting system Hamiltonian is a two-level atom in the presence of a uniform magnetic field of varying strength.

As evaluated in Appendix~\ref{Self-Commuting work distribution}, the probability of a measurement of the apparatus at some time $t$ giving the outcome $W$ is
\begin{align}
\mathcal{P}(W,t) 
&= \sum_n   \dfrac{\rho_S^{(n,n)}(t)}{\sqrt{\pi} \Sigma(t)}  e^{-\left( W - \int_{t_{p}}^{t_{m}} \dif{t} \, f(t) E_n(t) \right)^2/\Sigma^2(t)  } , \label{WorkDistribution}
\end{align}
where
\begin{align}
\rho^{(n,n)}_S(t) &=\bra{E_n} \rho_S(t) \ket{E_n}= \rho_{S}^{(n,n)}(t_{i}), \nn \\
\lambda\Sigma(t) &\ce \left(\frac{1}{\sigma_{p}^{2}} + \dfrac{\sigma_{p}^{2}(t-t_{p})^{2}}{m^{2}}\right)^{\frac{1}{2}}. 
 \nn
\end{align}
Note that the diagonal elements of state of the system do not evolve under $H_{S}(t)$. It is seen that the Gaussian factors appearing in Eq.~\eqref{WorkDistribution} disperse as $t$ increases in a nontrivial manner that depends on $m$ and $\sigma_p$. Moreover, $\sigma_{p}^{-1}$ quantifies how localized the initial apparatus state is in position space, so as $\sigma_p^{-1}$ decreases the measurement model approaches the ideal measurement limit.

Since the work distribution is simply a sum of Gaussians, the average work done on the system is 
\begin{align}
    \braket{W}_{\rm dist} = \sum_n   \rho_S^{(n,n)}(t_{i})  \int_{t_{p}}^{t_{m}} \dif{t} f(t) E_n(t) .  \label{distaverage}
\end{align}

In the ideal measurement limit, $f(t) \to \delta(t-t_f) - \delta(t-t_i)$, this expression reduces to the average work obtained from the first law using the freely evolving system state, $\rho_{S}$, so that $\Delta W_{\rm POVM} =0$ for arbitrary $m$ and $\sigma_p$. However, away from this limit $\Delta W_{\rm POVM}$ is in general non-zero. Moreover, as evaluated in Appendix~\ref{Self-Commuting work distribution}, we find that the quantity $\Delta W_{\rm int} $ defined in Eq.~\eqref{workint} vanishes independent of the shape of $f(t)$. Coupled with the fact that the diagonal elements of $\tilde{\rho}_{S}(t)$ are the same as those of $\rho_{S}(t)$ and Eq.~\eqref{firstlaw}, it follows that on average no heat transfer between the system and apparatus occurs. In general, this need not be the case, in particular when $[H_{S}(t),H_{S}(t')] \neq 0$ since $\Delta W_{\rm int}$ is non-zero and the diagonal elements of $\tilde{\rho}_{S}(t)$ are modified non-trivially.

Finally, the work distribution is seen to depend only on the diagonal elements in the energy eigenbasis of the system density matrix, which is a consequence of tracing out the system degrees of freedom in obtaining the reduced state of the apparatus. In the case of self-commuting system Hamiltonians, we find that the diagonal elements of the system density matrix do not evolve under $H_{S}(t)$ since the $\ket{E_{n}}$ remain eigenvectors for all $t \in \mathbb{R}^{+}$. We will see in Sec.~\ref{not self-commuting} that this no longer holds for the non-self-commuting case.

\subsection{Fluctuation relations}

Fluctuation relations are an important tool in statistical
mechanics because they relate equilibrium properties to measurable
non-equilibrium quantities. Generalizing classical fluctuation relations to the quantum regime has been the subject of much attention~\cite{perarnau-llobetQuantumSignaturesFluctuation2019,holmesCoherentFluctuationRelations2019,campisiColloquiumQuantumFluctuation2011,watanabeGeneralizedEnergyMeasurements2014}. Moreover, measurements of work fluctuations in quantum systems have been proposed and recently realized~\cite{Pekola2015CircuitQT,wuExperimentallyReducingQuantum2019, cerisolaUsingQuantumWork2017,perarnau-llobetCollectiveOperationsCan2018,perarnau-llobetNoGoTheoremCharacterization2017}.

Consider a system in contact with a heat bath of inverse temperature $\beta$, evolving under the system Hamiltonian $H_{S}(t)$. 
The Crooks relation connects the work distributions associated with the forward and backward protocols for an initial equilibrium thermal state, where the former corresponds to $\mathcal{P}_{F}(W,t)$ and the latter corresponds to $\mathcal{P}_{B}(-W,t_{m}-t)$ for $t\in [t_{p},t_{m}]$, are related~\cite{campisiColloquiumQuantumFluctuation2011,crooksEntropyProductionFluctuation1999}, and can be stated as 
\begin{align}
    \mathcal{P}_{F}(W) = \mathcal{P}_{B}(-W) e^{\beta(W-\Delta F)}, \nn
\end{align}
where 
$\Delta F$ is the change in the equilibrium free energy of the system, defined by $\Delta F \ce -\frac{1}{\beta}
\ln{\tfrac{Z(t_{f})}{Z(t_{i})}}$, and $Z(t) \ce \tr e^{-\beta H_S(t)}$ is the partition function of the system at time $t$. 
If we consider an initial thermal state, $\tilde{\rho}_{S}(t_{i})=\frac{e^{-\beta H_{S}(t_{i})}}{Z(t_{i})}$,
the work done on the system obeys $W= - \Delta F$ in the ideal measurement limit. Then, the Crooks relation simply states that $ \frac{\mathcal{P}_{F}(W)}{\mathcal{P}_{B}(-W)} =  e^{2\beta W}$.
Using Eq.~\eqref{WorkDistribution}, we obtain the work distribution associated to this thermal state
\begin{align}
\mathcal{P}_{F}(W,t)&= \frac{1}{Z(t_{i})\sqrt{\pi}   \Sigma(t)} \nn \\
&\times \sum_n  e^ {- \beta E_n(t_i) } e^{ -\frac{\left(W - \int_{t_p}^{t_m} \dif{t} \lambda f(t) E_n (t) \right)^2 }{  \Sigma(t)^2}} \label{ThermalStateDistribution} .
\end{align}
Using this work distribution, which takes into account the act of measuring the system on which work is being performed, we arrive at a modified Crooks relation specific to our measurement model: 
\begin{widetext}
\begin{align}
 \dfrac{\mathcal{P}_{F}(W,t)}{\mathcal{P}_{B}(-W,t_m -t)}=  \dfrac{\Sigma(t_{m}-t)Z(t_f)}{\Sigma(t)Z(t_{i})} \dfrac{ \sum_n  e^ {-
  \beta E_n(t_{i}) } e^{ -\left(W - \int_{t_p}^{t_m} \dif{t} \,  f(t) E_n
  (t) \right)^2 /   \Sigma(t)^2}}{\sum_n  e^
    {- \beta E_n(t_{f}) } e^{ -\left(W + \int_{t_{m}}^{t_{p}} \dif{t} \,
     f(t_{m}-t) E_n (t_{m}-t) \right)^2 /   \Sigma(t_{m}-t)^2}},
    \label{Crooksmodified}
\end{align}
\end{widetext}
where we have parameterized the forward protocol with $t$ for $t_{p}\leq t\leq t_{m}$ and the
backward protocol with $t_{m}-t$ for $t_{p}\leq T\leq t_{m}$.  Equation~\eqref{Crooksmodified} constitutes a generalization of the standard Crooks relation when the ancilla-assisted measurement protocol is used to define work and finite measurement interactions times and dispersion effects in the measuring apparatus are taken into account. Note that by taking the ideal measurement limit discussed under Eq.~\eqref{Effect}, we reproduce the Crooks relation for equilibrium states.

The Jarzynski equality is another important fluctuation relation that governs systems away from equilibrium~\cite{campisiColloquiumQuantumFluctuation2011,JarzynskiEquality1996}, which can be derived straightforwardly from the Crooks relation 
\begin{align}
    \Braket{e^{-\beta W}} = \int \dif{W} \mathcal{P}_{F}(W)e^{-\beta W} = e^{-\beta \Delta F}= \dfrac{Z(t_{f})}{Z(t_{i})}. \label{jarz}
\end{align}
Moreover, using Jensen's inequality, $\Braket{e^{-\beta W}} \geq e^{-\beta \Braket{W}}$, the statement of the 
second law of thermodynamics follows
\begin{align}
    \Braket{W} \geq \Delta F.
    \label{secondlaw}
\end{align}

By using the work distribution in Eq.~\eqref{ThermalStateDistribution}, we can calculate the
exponentiated average work at the time of measurement of the apparatus, $t_{m}$, and use that to arrive at a modified Jarzynski equality
\begin{align}
    \Braket{e^{-\beta W}}_{\rm dist}
= \frac{e^{\frac{\beta^{2} \Sigma^{2}(t_{m})}{4}} }{Z(t_i)}      \sum_n  e^{    - \beta \left( E_n(t_i) +  \int_{t_p}^{t_m} \dif{t} \, f(t) E_n (t) \right)  }.  \nn 
\end{align}
Upon taking the ideal measurement limit, the modified
Jarzynski equality reduces to the standard Jarzynski equality in Eq.~\eqref{jarz}. These corrections are similar to those found in Ref.~\cite{talknerAspectsQWork2016, solinasvariance2017}, especially the constant exponential correction $e^{\beta^{2} \Sigma^{2}(t_{m})/4}$, except that our measurement model takes into account the finite duration of the interaction and the mass of the detector through the dependence on $\Sigma$ and not just $\sigma_{x}$. Moreover, the remaining corrections depend on how the apparatus samples the energy of the system through the $f(t)$ term appearing in $H_{\rm int}$.

For an equilibrium state of the system at inverse temperature $\beta$, Eq.~\eqref{distaverage} simplifies to the following
\begin{align}
    \Braket{W}_{\rm dist} = \frac{1}{Z(t_{i})} \sum_{n}  e^{-\beta E_{n}(t_{i})} \int_{t_{p}}^{t_{m}} \dif{t}  f(t) E_{n}(t)   \nn 
\end{align}

Using the same reasoning that led to Eq.~\eqref{secondlaw}, we arrive at a statement of the second law 
of thermodynamics with respect to our measurement model 
\begin{align}
    \Braket{W}_{\rm dist} &\geq 
    -\frac{1}{\beta} \ln{\Bigg(\frac{1}{Z(t_i )}\sum_{n}e^{-\beta(E_{n}(t_{i})+ \int_{t_{p}}^{t_{m}} \dif{t}  f(t) E_{n}(t))}\Bigg)} \nn \\ &\quad -\frac{\beta \Sigma^{2}(t_{m})}{4} . \nn
\end{align}

It is seen that the finite resolution of the measurement modifies the expression of the second law in a way that is dependent on the temperature of the system and the measurement model parameters. In the ideal measurement limit, the first term in the above expression reduces to $\Delta F$ while the second term goes to zero, thus reproducing the expression in Eq.~\eqref{secondlaw}. We find that there is a constant correction proportional to the product of $\beta$ and the square of the width of the work distribution at $t_m$. For sufficiently low temperatures, this constant correction may remain non-zero even in the ideal measurement limit. 

%========================================
%========================================
\section{The work done on a two-level atom by a changing magnetic field}
\label{not self-commuting}
We now consider the case in which the system Hamiltonian $H_S(t)$ does not commute with itself at different times, $[H_{S}(t),H_{S}(t')] \neq 0$. As an example of such a scenario, we consider a two-level atom $\mathcal{H}_S \simeq \mathbb{C}^2$ in the presence of a magnetic field that changes in strength and direction between the times $t_p$ to $t_m$,
\begin{align}
H_S(t) = \mu \vec{B}(t) \cdot \vec{\sigma}, \nn 
\end{align}
where $\mu$ is the magnetic moment of the atom, $\vec{B}(t) = B(t) \hat{n}(t)$ is the magnetic field vector and $\vec{\sigma} = (\sigma_x, \sigma_y, \sigma_z)$ is the Pauli vector. For simplicity, we suppose that the magnetic field is rotating around the $z$-axis at a polar angle $\theta$ so that in the basis furnished by the eigenstates of the $\sigma_z$ operator, the system Hamiltonian takes the form
\begin{align}
H_S(t) = \mu B(t) \left[  \cos \omega t  \sin \theta \sigma_x +  \sin \omega t   \sin \theta \sigma_y +   \cos \theta \sigma_z \right]. \nn
\end{align}
This Hamiltonian does not commute with itself at different times,
$[H_{S}(t),H_{S}(t')] \propto \sin{\theta}$, unless $\theta$ is an integer multiple of $\pi$ in which case the results developed in Sec.~\ref{subsectionselfcommutingtheory} apply. Thus, we will use the parameter $\theta$ as a measure of the non-self-commutativity of $H_{S}(t)$.

Suppose that the system and measuring apparatus are prepared at time $t_p$ in a product state $\ket{\psi_A(t_p)} \ket{\psi_S(t_p)}$, where $\ket{\psi_A(t_p)}$ is given in Eq.~\eqref{Astatepgaus} and the initial state of the system at the time of the first sampling $t_i$ is
\begin{align}
\ket{\psi_S(t_i)} = \alpha \ket{0} + \beta \ket{1}, \nn 
\end{align}
where $\alpha$ and $\beta$ are complex numbers such that $\abs{\alpha}^2 + \abs{\beta}^2 = 1$. To properly compare the effects of the parameters defining the measurement model ($\Delta$, $m$, $\sigma_{p}$, $\lambda$, $\omega$) with an the ideal measurement as described in Sec.~\ref{secideal}, $\alpha$ and $\beta$ are chosen such that if the system were to evolve under $H_S(t)$ alone, then at $t_i$ the system would be in the state
\begin{align}
\mathcal{T} e^{-i \int_{t_p}^{t_i} \dif{t} \, H_S(t)}  \ket{\psi_S(t_p)} =   \alpha \ket{0} + \beta \ket{1}. \nn
\end{align}

Generally, we can expand the joint state of both the apparatus and the system at a time $t$ as
\begin{align}
    \ket{\psi(t)}= \sum_{n \in \{0,1\}} \int \dif{p} \, c_{n}(t,p) \ket{n} \ket{p},
    \label{jointstate} 
\end{align}
where we have expanded the system's state in the $\sigma_{z}$-basis. The coefficient
functions $c_{n}(t,p)$ can be determined by substituting Eq.~\eqref{jointstate} in the Schr\"{o}dinger 
equation of the Hamiltonian  in Eq.~\eqref{totalH}. We arrive at two coupled 
differential equations 
\begin{align}
    i\dot{c}_{j}(t,p) = \frac{p^{2}}{2 m} c_{j}(t,p) + \left[1 + \lambda f(t) p \right] \sum_{k} c_{k}(t,p)
    H_{jk}(t),\label{maindifeq}
\end{align}
 where $j,k \in \{0,1\}$ and we have defined $H_{jk}(t) = \braket{j|H_{S}(t)|k}$.  The work distribution is obtained from the diagonal entries in the position basis of the reduced apparatus density matrix. In
this case, the work distribution is given by
    \begin{align}
\mathcal{P}(W,t) &= \frac{1}{2 \pi } \sum_{n}\int \dif{p}\dif{p'} \, c_n(t,p)  c_n^*(t,p')  e^{iW(p-p')}.  \nn \\
&=  \frac{1}{2 \pi } \sum_{n}\abs{\int \dif{p} \, c_n(t,p)   e^{i\lambda Wp}}^2. \nn
\end{align}

 \begin{figure}[t]
     \centering
     \includegraphics[width=0.45\textwidth]{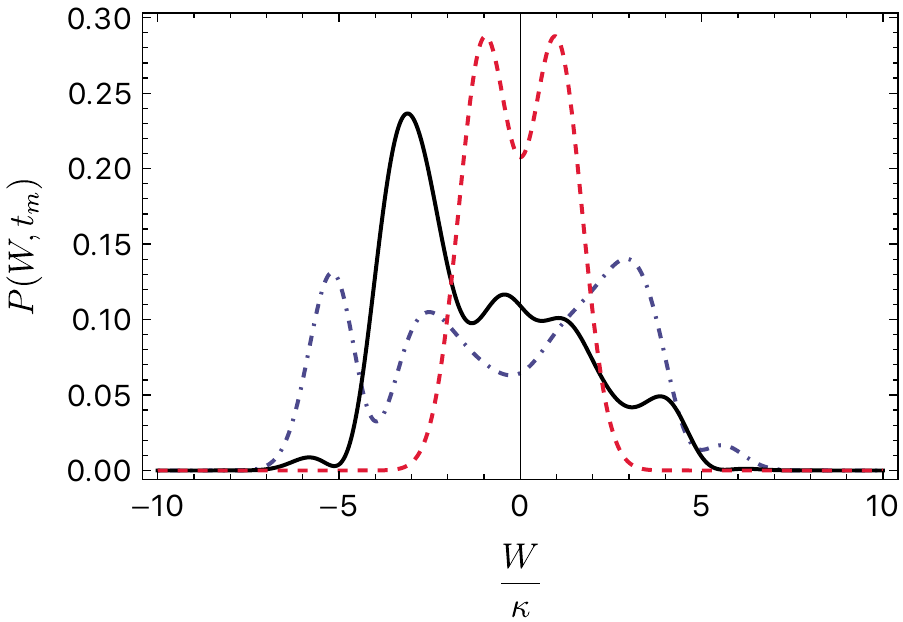}
      \includegraphics{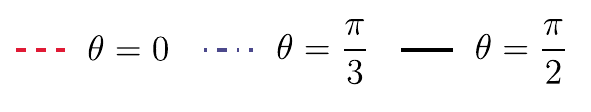}
     \caption{For a system prepared in an equally weighted superposition of the two energy
     eigenstates, which corresponds to $\alpha=\tfrac{1}{\sqrt{2}}$ at the time of the first measurement $\kappa t_{i}=2$, we show the work distribution at $\kappa t_{m}=4$ for different values of $\theta$.The mass of the measuring apparatus and the duration of the interaction are taken to be $\frac{\kappa m}{\sigma_{p}^{2}}=1000$ and $\kappa \Delta = 0.2$.}
     \label{thetacomparison}
 \end{figure} 

\begin{figure}[ht]
     \centering
     \includegraphics[width=0.45\textwidth]{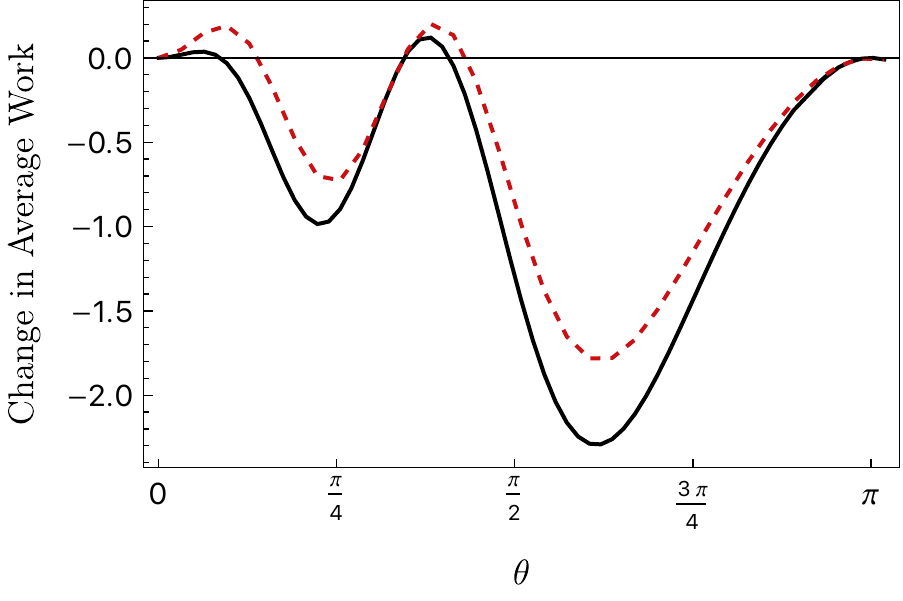}\\
      \includegraphics[width=0.25\textwidth]{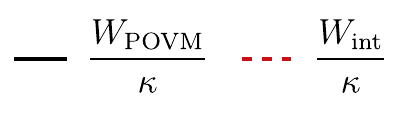} 
     \caption{For an initially excited two-level system, corresponding to $\alpha=0$ at the time of the first measurement $\kappa t_{i}=2$, the differences in average work $\frac{\Delta W_{\rm int}}{\kappa}$ and $\frac{\Delta W_{\rm POVM}}{\kappa}$, as defined in Eqs.~\eqref{workint} and \eqref{davgworkpovm}, are plotted at the time of measurement $\kappa t_{m}=4$ for $\theta \in [0,\pi]$ . In general, these differences are non-zero, illustrating a discrepancy between the average estimated work and modification due to the act of measuring the work distribution with the average work defined in Eq.~\eqref{sswork} corresponding to the work done in the absence of measurement. The mass of the measuring apparatus and the duration of the interaction are taken to be $\frac{\kappa m}{\sigma_{p}^{2}}=1000$ and $\kappa \Delta = 0.2$. } 
     \label{avgworkplot}
 \end{figure} 
The coupled differential equations in Eq.~\eqref{maindifeq} can be solved
numerically and their solutions used to arrive at a work distribution
specific to the measurement model. 

To illustrate measurement interaction effects on the ancilla-assisted measurement protocol in a concrete example, consider $B(t)= \gamma t$, where $\gamma$ has units of magnetic field times energy. The overall factor entering the system Hamiltonian is $\kappa^{2} := \mu \gamma$, where $\kappa$ has units of energy or inverse time. This constant dictates the system dynamics, so we express our relevant physical parameters in terms of it. In the case where a physical parameter has momentum dependence, we express it in terms of $\sigma_{p}$. Moreover, suppose that $g(t) = \frac{1}{\sqrt{\pi \Delta^2 }}e^{-t^2/ \Delta^2}$ with the interpretation that the duration of the interaction between the  apparatus and system is on the order of $\kappa \Delta$. To characterize this process, plotted in Fig.~\ref{thetacomparison} is the estimated work distribution and in Fig.~\ref{thetacomparison} the discrepancy between the average estimated work and the average work defined in Eq.~\eqref{sswork} as a function of $\theta$. In all figures, the first interaction occurs at $\kappa t_{i}= 2$, the second interaction at $ \kappa t_{f}= 3$, and the measurement of the work distribution at $ \kappa t_{m}= 4$.

From Fig.~\ref{thetacomparison}, we note that the general structure of the work distribution in Eq.~\eqref{WorkDistribution} as a weighted sum of Gaussian functions centered at different work values remains just as in the self-commuting case. This is because the system Hamiltonian at $t_1$ has energy-eigenvalues $\{+\mu B(t_1),-\mu B(t_1) \}$ and which evolve to $\{+\mu B(t_2),-\mu B(t_2) \}$ at time $t_2$, from which it is seen that in general, there will be four possible energy exchange modes for the two-level system. Given our choice of parameters, these modes correspond to $\frac{W}{\kappa} \in \{\pm 1, \pm 5\}$. Numerically, we see that as $\theta$ increases, the exterior peaks (those at $\pm 5$) of the work distribution gain non-trivial amplitudes in contrast to the $\theta =0$ case. Recall that $[H_{S}(t), H_{S}(t')] \propto \sin \theta$, and so variation in $\theta$ modifies the evolution of the state of the system, leading to different probability amplitudes for the expected modes of energy exchange. In addition, the location of the peaks of the work distributions is displaced non-trivially relative to the classically expected locations as $\theta$ is varied.

Figure~\ref{avgworkplot} is a plot of $\Delta W_{\rm POVM}$ and $\Delta W_{\rm int}$, defined respectively in Eqs.~\eqref{workint} and \eqref{davgworkpovm}, which quantify the discrepancy in estimated average work and the additional work done by the act of measuring. It is observed that in general these quantities are non-zero, reflecting model parameters different from the ideal measurement limit; in the ideal limit, $\Delta W_{\rm POVM},\Delta W_{\rm int} \to 0$ as expected. We emphasize that this deviation in the average work can be quite large.  A similar analysis of the average heat exchanged between the measuring apparatus and system leads to the conclusion that $\Delta Q_{\rm int}$ as defined in Sec.~\ref{thermocons} is generically non-zero when $\theta \neq 0$. This can be seen from the non-trivial modification to the reduced state of the system
\begin{align}
    \tilde{\rho}_{S}(t) = \sum_{n,m} \int \dif{p} c_{n}(t,p) c^{*}_{m}(t,p) \ket{n} \bra{m}, \nn
\end{align}
constructed from the solutions to Eq.~\eqref{maindifeq}, on account of the system's interaction with the apparatus.

\begin{figure}[ht]
     \centering
     \includegraphics[width=0.45\textwidth]{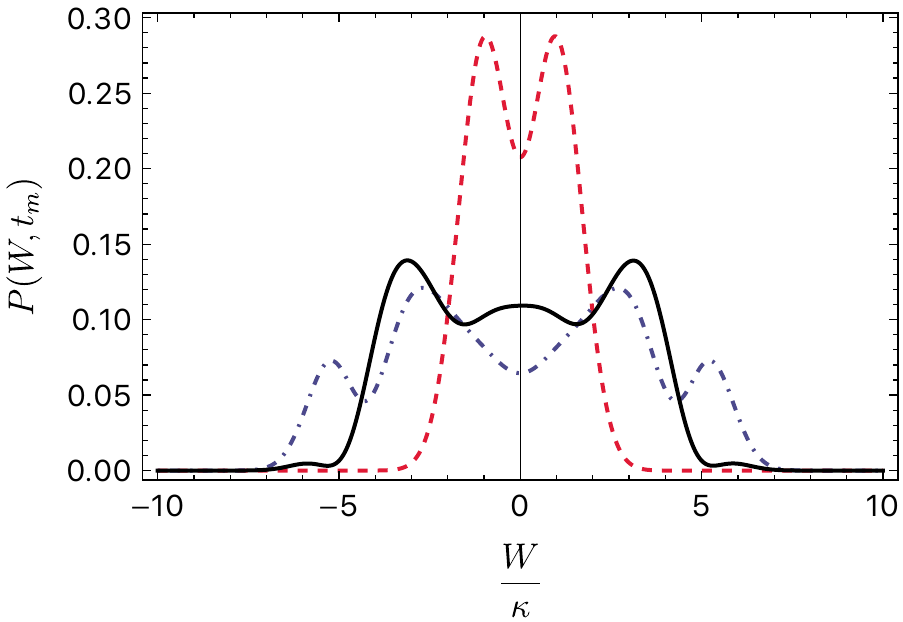}\\
      \includegraphics[width=0.25\textwidth]{tcomplegend.pdf} 
     \caption{For an initial equally-weighted classical mixture, corresponding to $\rho_{S}(t_{i})= \frac{1}{2} \left( \ket{0}\bra{0} + \ket{1}\bra{1}\right)$ at the time of the first measurement $\kappa t_{i}=2$, we show the work distribution at $\kappa t_{m}=4$ for the same values of $\theta$ displayed in Fig.~\ref{thetacomparison}. The mass of the measuring apparatus and the duration of the interaction are taken to be $\frac{\kappa m}{\sigma_{p}^{2}}=1000$ and $\kappa \Delta = 0.2$. } 
     \label{mixplot}
 \end{figure}

A key difference between the two-point measurement scheme and the POVM approach to estimating the work distribution is that the latter keeps a coherent record of the energy of the system of interest. The former, where projective measurements are employed, projects the system into different energy eigenstates so that any coherence in the initial state is gone while the time-dependent Hamiltonian is doing work. In Fig.~\ref{mixplot}, we plot the work distribution for an initial system state that is in a classical equally-weighted mixture. The behavior of the work distribution is expected to be different from that in Fig.~\ref{thetacomparison}, which is for an equally-weighted \textit{superposition} initial system state. 

%========================================
%========================================
%========================================
%========================================
\section{Conclusion}
\label{Conclusion}
The ancilla-assisted protocol for measuring work distributions was generalized to account for dispersion and finite resolution effects of the measuring apparatus used to extract the work distribution. An explicit measurement model was considered that replicates the statistics of the probability distribution associated with the work done on the system. Two regimes were explored, one in which the system Hamiltonian self-commutes with itself at different times, and another in which it does not. The former admits an analytic expression for the work distribution, which we obtain, while the latter does not but was explored numerically via an example of a two-level in a time-dependent magnetic field. Corrections to the Crooks relation and Jarzynski equality were shown to manifest on account of the finite resolution of the measuring apparatus, which are expected to manifest in any realistic measurement of work distributions.
\begin{acknowledgments} 
We thank Nicolai Friis and Nicole Yunger Halpern for their expertise and careful revision of a draft of this paper. We'd also like to thank an anonymous referee for their helpful comments. This work was supported in part by the Paul K. Richter and Evalyn E. Cook Richter Memorial Fund, a Kaminsky Undergraduate Research Award, the Dartmouth College Society of Fellows, and Saint Anselm College. 
\end{acknowledgments}
%\IEEEtriggeratref{~}
%========================================
%========================================
\appendix
\onecolumngrid
\section{Self-Commuting work distribution}
\label{Self-Commuting work distribution} 
We consider the case where the Hamiltonian commutes with itself at different times, and so the 
time-evolution operator governing the evolution of the state does not need to be time-ordered; expanding this operator in the energy eigenbasis of the system Hamiltonian yields
   \begin{align}
U_S(t)=  \sum_ne^{-i \int_{t_p}^{t_m} \dif{t} \, E_n(t)} \ket{E_n} \!\bra{E_n}, \nn 
\end{align} 
The evolution of the joint system is governed by the following unitary operator
\begin{align}
    U(t) = U_{SA}(t)[U_{S}(t)\otimes U_{A}(t)] \nn 
\end{align}
Because the system Hamiltonian commutes with itself at different times and, because of the
form of the interaction Hamiltonian, it follows that all three terms contribution to the total Hamiltonian in Eq.~\eqref{totalH} commute among themselves. Evolving the initial state of the system and apparatus yields 
\begin{align}
\rho_{SA}(t)
 &=
\sum_{m,n} \int \dif{p} \dif{p'} \,  e^{-i p \int_{t_p}^{t_m} \dif{t} \, \lambda f(t) E_n(t)}    \rho^{(m,n)}_S(t) e^{i p' \int_{t_p}^{t_m} \dif{t}  \, \lambda f(t)E_{m}(t) } 
\rho_{A}^{(p,p')}(t)   \ket{E_m}\!\bra{E_n} \otimes  \ket{p} \! \bra{p'},  \label{JointStateappendix}
\end{align}
where the free-evolution of the reduced state of the apparatus state appearing above is
\begin{align}
  \rho^{(p,p')}_{A}(t)  = \psi_{A}(p,t)\psi^{*}_{A}(p',t) \qquad \text{where} \qquad \psi_{A}(p,t) = \frac{1}{\sqrt{\pi^{1/2} \sigma_p}}
 e^{-\left(\frac{1}{\sigma_p^2}+\frac{i(t-t_{p}) }{m }\right) \frac{p^2}{2}}. \nn 
\end{align}
By construction, the amount of work done on the system by $H_S(t)$ is encoded in the position degree
of freedom of the measuring apparatus. The reduced state of the apparatus is obtained by tracing over the system Hilbert space,
\begin{align}
\tilde{\rho}_A(t) 
&=\sum_{n} \int \dif{p} \dif{p'} \,  e^{{-i(p-p') \left( \int_{t_p}^{t_m} \dif{t} \, \lambda f(t) E_n(t) \right)}} \rho^{(n,n)}_S(t) \rho_{A}^{(p,p')}(t)  \ket{p} \! \bra{p'}, \nn
\end{align}
This state may be expressed in the position basis as
\begin{align}
\tilde{\rho}_A(t) 
&= \frac{1}{2 \pi} \sum_{n}\int  \dif{x} \dif{x'} \, \dif{p} \dif{p'} \,  e^{i  (xp-x'p') }e^{-i (p-p') \int_{t_p}^{t_m} \dif{t} \, \lambda f(t) E_n(t)  }    \rho^{(p,p')}_{A}(t) \rho_S^{(n,n)}(t)  \ket{x} \! \bra{x'},  \nn
\end{align}
Finally, the work distribution of the system is given by the diagonal elements of the $\tilde{\rho}_A(t)$ in the position basis
\begin{align}
\mathcal{P}(W,t)  
&= \frac{1}{2 \pi} \sum_n \rho_S^{(n,n)}(t)\int  \dif{p} \dif{p'} \,   e^{{i\lambda(p-p') \left( W - \int_{t_p}^{t_m} \dif{t} \,  f(t) E_n(t) \right)}} \rho^{(p,p')}_{A}(t) 
\label{workdistappendix} 
\end{align}
Defining $a\ce \frac{1}{\sigma_p^2}+i \frac{(t-t_{p}) }{m }$, consider the integral appearing in Eq.~\eqref{workdistappendix} 
\begin{align}
    &\frac{1}{2\pi} \frac{1}{\sqrt{\pi}\sigma_{p}} \int  \dif{p} \dif{p'} \,   e^{{i\lambda(p-p') \left( W - \int_{t_p}^{t_m} \dif{t} \,   f(t) E_n(t) \right)}} e^{- \frac{a p^2 + \bar{a} p'^{2}}{2}}  \nn \\
    &\qquad \qquad = \frac{1}{2 \pi^{\frac{3}{2}} \sigma_{p}} \int \dif{p} e^{i\lambda p\left( W - \int_{t_p}^{t_m} \dif{t} \,   f(t) E_n(t) \right)} e^{-\frac{ap^{2}}{2}}\int \dif{p'} e^{-i\lambda p'\left( W - \int_{t_p}^{t_m} \dif{t} \, a  f(t) E_n(t) \right)} e^{-\frac{\bar{a }p'^{2}}{2}} \nn \\
    &\qquad \qquad= \frac{1}{\sqrt{\pi} |a| \sigma_{p}} e^{-\lambda\frac{\left( a^{-1}+\bar{a}^{-1}\right)}{2}\left( W - \int_{t_p}^{t_m} \dif{t} \,   f(t) E_n(t) \right)^{2} } \nn \\
    &\qquad \qquad= \frac{1}{\sqrt{\pi} |a|\sigma_{p} } e^{-\lambda\left( \frac{\Re[a]}{|a|^{2}}\right)\left( W - \int_{t_p}^{t_m} \dif{t} \,   f(t) E_n(t) \right)^{2} } \nn \\
    &\qquad \qquad= \frac{1}{\sqrt{\pi} \Sigma(t) } e^{-\left( W - \int_{t_{p}}^{t_{m}} \dif{t} \, f(t) E_n(t) \right)^2/\Sigma^2(t)  },
    \label{derivation}
\end{align}
where we have defined $\lambda\Sigma(t) \ce \sigma_{p} |a|$. Substituting Eq.~\eqref{derivation} into Eq.~\eqref{workdistappendix} we find
\begin{align}
    \mathcal{P}(W,t) &= 
    \sum_n   \dfrac{\rho_S^{(n,n)}(t)}{\sqrt{\pi} \Sigma(t)}  e^{-\left( W - \int_{t_{p}}^{t_{m}} \dif{t} \, f(t) E_n(t) \right)^2/\Sigma^2(t)  }.
\end{align}
where $\rho_{S}^{(n,n)}$ are the diagonal elements of the system state subject to evolution under $H_{S}(t)$ alone.

Conversely, we can obtain the reduced system state $\tilde{\rho}_S(t)$ by tracing out the apparatus in Eq.~\eqref{JointStateappendix}:
\begin{align}
\tilde{\rho}_S(t)  &= \sum_{m,n} \int \dif{p} \,  e^{-i p \int_{t_p}^{t_m} \dif{t} \, \lambda f(t) \left(E_n(t)-E_m(t)\right)  }    \rho^{(m,n)}_{S}(t) \rho_{A}^{(p,p)}(t) \ket{E_m}\!\bra{E_n}  \nn \\
&=\frac{1}{\sqrt{\pi} \sigma_{p}}\sum_{m,n} \rho^{(m,n)}_{S}(t) \int \dif{p} \,  e^{- \frac{p^{2}}{\sigma_{p}^{2}}-i p \int_{t_p}^{t_m} \dif{t} \, \lambda f(t) \left(E_n(t)-E_m(t)\right)}  \ket{E_m}\!\bra{E_n}  \nn \\
&= \sum_{m,n} \rho^{(m,n)}_{S}(t) e^{-\frac{1}{{4\sigma_{x}^{2}}}\left(\int_{t_p}^{t_m} \dif{t} \, \lambda f(t) \left[E_n(t)-E_m(t)\right]\right)^{2} }\ket{E_m}\!\bra{E_n}.
\label{reducedStateappendix}
\end{align}

Using the definition in Eq.~\eqref{Defworktilde} and the above expressions, an analytic expression of the average work is obtained:
\begin{align}
    \braket{\tilde{W}}_{S} &= \int \dif{t} \tr \left[ \dot{H}_{S} \tilde{\rho}_{S}(t)\right] \nn \\
    &=\sum_{m,n} \int \dif{t} \rho^{(m,n)}_{S}(t) e^{-\frac{1}{{4\sigma_{x}^{2}}}\left(\int_{t_p}^{t_m} \dif{t'} \, \lambda f(t') \left[E_n(t')-E_m(t')\right]\right)^{2} }\tr \left[ \dot{H}_{S} \ket{E_m}\!\bra{E_n} \right] \nn \\ 
    &= \sum_{n} \rho^{(n,n)}_{S}(t_i ) \int \dif{t} \dot{E}_{n}(t) \nn \\
    &= \braket{W}_{S}. \nn
\end{align}
This means that $\Delta W_{\rm int}=0$ for self-commuting system Hamiltonians, in contrast to $\Delta W_{\rm POVM}$,
\begin{align}
   \Delta W_{\rm POVM} = \sum_{n} \rho_{S}^{(n,n)} \left(\int_{t_{i}}^{t_{f}} \dif{t}\lambda f(t) E_{n}(t) - \left[E_{n}(t_{f}) - E_{n}(t_{i})  \right] \right),  \nn
\end{align}
which only vanishes in the ideal measurement limit.
\end{document}